# Letter

# Novel magnetic properties of graphene: Presence of both ferromagnetic and antiferromagnetic features and other aspects

**H. S. S. Ramakrishna Matte, K. S. Subrahmanyam and C. N. R. Rao***

International Centre for Materials Science, New Chemistry Unit and CSIR Centre of Excellence in Chemistry, Jawaharlal Nehru Centre for Advanced Scientific Research, Jakkur P. O., Bangalore 560 064 (India).

## Abstract:

Investigations of the magnetic properties of graphenes prepared by different methods reveal that dominant ferromagnetic interactions coexist along with antiferromagnetic interactions in all the samples. Thus, all the graphene samples exhibit room-temperature magnetic hysteresis. The magnetic properties depend on the number of layers and the sample area, small values of both favoring larger magnetization. Molecular charge-transfer affects the magnetic properties of graphene, interaction with a donor molecule such as tetrathiafulvalene having greater effect than an electron-withdrawing molecule such as tetracyanoethylene.

* For correspondence: cnrrao@jncasr.ac.in, Fax: (+91) 80-2208 2760

It was pointed sometime ago that edges in graphene ribbons play a crucial role in determining the electronic structure.[1] There have been a few studies on the properties of nanographite particles and ribbons which demonstrate the importance of the edge states arising from the nonbonding electrons.[2-4] Nanographite particles are reported to exhibit unusual magnetic properties including spin-glass behavior and magnetic switching phenomena.[2-4] Bi-layer graphene is predicted to be ferromagnetic.[5] Hydrogenated nanographite is also predicted to show spontaneous magnetism.[6] Magnetic properties of nanographite have been reviewed by Enoki et al.[7-9] and the main message is that edge states as well as adsorbed or intercalated species affect the magnetic properties. By carrying out first-principles density functional calculations, Lee et al.[10] have shown the existence of a ferromagnetically ordered ground state in the zig-zag edges of graphene. Zig-zag edges longer than 3-4 repeat units are predicted to be magnetic, irrespective of whether the edges are regular or irregular.[11] We have carried out an experimental study of the magnetic properties of graphene samples prepared by different methods. Since magnetism of graphene is due to edge states, we considered it worthwhile to investigate the effect of electron donor and acceptor molecules on the magnetic properties. With this purpose, we have measured the effect of interaction of graphene with tetrathiafulvalane (TTF) and tetracyanoethylene (TCNE). These molecules are known to significantly affect the electronic properties and Raman spectra of graphene because of charge-transfer.[12] Significantly, we find that all the graphene samples show evidence for ferromagnetism with well-defined hysteresis, along with antiferromagnetic features. Electron-donating TTF markedly affects the magnetic properties of graphene.

Graphene samples were prepared by thermal exfoliation of graphitic oxide (EG),[13,14] conversion of nanodiamond (DG)[2,13,14] and arc evaporation of graphite in hydrogen (HG).[15] We also prepared a graphene sample (EG-H) by reducing graphene oxide with hydrazine hydrate.[16] The samples were characterized using transmission electron microscopy (TEM), atomic force microscopy (AFM), Raman spectroscopy and Brunauer-Emmett-Teller (BET) surface areas. Magnetic measurements were performed with a vibrating sample magnetometer (VSM) in physical property measuring system (Quantum Design, USA). The effect of adsorption of 0.01M and 0.05 M benzene solutions of TTF and TCNE on the magnetic properties of HG was also studied. Electron paramagnetic resonance (EPR) spectra were recorded using a Bruker EMX X-band continuous wave (CW) EPR spectrometer.

In Figure 1, we show typical AFM data and Raman spectra in the case of EG and HG. Unlike the HG sample, the EG sample consists of a larger number of layers. From the AFM measurements, the average number of graphene layers was estimated to be 6-7, 4-5 and 2-3 repectively in EG, DG and HG. Raman spectra show the D-, G-, and 2D bands around 1320, 1570 and 2640 $cm^{-1}$ and the relative intensity of the defect-related D-band is much higher in HG. The BET surface areas[17] of the samples were 826 $m^2/g$, 781 $m^2/g$ and 680 $m^2/g$ respectively. The areas of the graphene flakes in the EG, DG and HG samples were $65\pm25\times10^3$ $nm^2$, $40\pm20\times10^3$ $nm^2$ and $25\pm15\times10^3$ $nm^2$ respectively. Of the three graphene samples, HG had the smallest number of graphene layers and the smallest area. The EG-H sample appears to possess even a smaller number of layers close to unity.[16]

In Figure 2, we show the temperature-dependence of magnetization of EG, DG and HG measured at 500 Oe. All the samples show divergence between the field-cooled (FC) and zero-field-cooled (ZFC) data, starting around 300 K. The divergence nearly disappears on the application of 1 T as can be seen from the insets in Figure 2. EG-H also shows a behavior similar EG, DG and HG. Divergence between the FC and ZFC data in the graphene samples is comparable to that in magnetically frustrated systems such as spin-glasses and superparamagnetic materials. The Curie-Weiss temperatures obtained from the high-temperature inverse susceptibility data were negative in all these samples, indicating the presence of antiferromagnetic interactions. Interestingly, we observe well-defined maxima in the magnetization at low temperatures, the maxima becoming prominent in the data recorded at 1 T (see insets in Figure 2). Such magnetic anomalies are found when antiferromagnetic correlations compete with ferromagnetic (FM) order. Application of high fields aligns the FM clusters and decreases the divergence between FC and ZFC data as indeed observed. It is possible that the data correspond to percolation type of situation, where in different types of magnetic states coexist. The FM clusters in such a case would not be associated with a well-defined global ferromagnetic transition temperature. This behavior is similar to that of microporus carbon and some members of the rare earth manganite family, $Ln_{1-x}A_xMnO_3$ (Ln=rare-earth, A=alkaline earth).[18] Recent theoretical calculations do indeed predict the presence of antiferromagnetic states in the sheets and ferromagnetic states at the edges of graphene.[19]

All the graphene samples show magnetic hysteresis at room temperature (Figure 3). Of the three samples, HG shows the best hysteretic features with saturation. While DG also shows saturation, the value of saturation magnetization, $M_S$, is, however, low compared to HG. In Table 1, we list the values of $M_S$, remnant magnetization, $M_R$, and the Curie-Weiss temperature, $\theta_p$, along with value of magnetization at 100 K (at 3000 Oe) for the three graphene samples studied by us. We see that $\theta_p$, $M_R$ and $M_S$ are highest in case of HG which also shows higher value of magnetization than the other samples at all temperatures. We have plotted the values of the various magnetic properties of the samples in Figure 4(a) to demonstrate how the properties vary as HG > DG > EG. In Figure 4 (b) we have shown the variation of the average area and the number of layers in these three samples. It is noteworthy that both the area and the number of layers decrease in the order of EG> DG> HG. It is likely that the edge effects would be greater in samples with smaller number of layers as well as small areas. Since we completed research on the magnetic properties of graphene we have noticed a report of ferromagnetism in a graphene sample by Wang et al.[20] These workers prepared graphene by partially reducing graphene oxide with hydrazine and annealing the samples at different temperatures in an argon atmosphere. The value of $M_S$ found by us is much larger than that reported by Wang et al. Furthermore, saturation is attained above 1500 Oe in all our samples. The EG-H sample prepared by us exhibits very large magnetization, much larger than HG.[21] It must be noted that in all our synthesis, we avoided transition metal impurities. The unusual magnetic properties reported here are, therefore, intrinsic to the graphene samples.

Since we find the presence of AFM interactions as well as magnetic hysteresis in our graphene samples, a behavior somewhat like that of frustrated magnetic systems, we carried out ac susceptibility measurements on HG and DG samples in the frequency range from 97 to 9997 Hz. We did not find any frequency-dependent features in the ac susceptibility data in the 3-300 K range. This suggests that observed ferromagnetism is not due to spin-glass behavior.

We have carried out EPR investigations on EG, DG and HG samples in the temperature ranging from 2.5 K to 300 K. We have observed a signal with a line-width of $\Delta H \approx 0.7\text{-}2.9$ mT with a g- value is in the 2.006-2.013 range. The small value of the line-width and the small deviation in the g value from the free-electron value suggest that the spins do not originate from transition-metal impurities but from only carbon-inherited spin species in the graphene sheets. The temperature variation of the EPR intensity is consistent with the magnetization data showing a marked increase at low temperatures. Spin-density measurements were carried on EG, DG and HG samples taking copper sulfate as a reference, gave values of $2.86 \times 10^{12}$, $1.48 \times 10^{13}$ and $2.46 \times 10^{14}$ respectively.

Adsorption of benzene solutions of TTF and TCNE has a profound effect on the magnetic properties of graphene. In Figure 5, we show typical results on the effect of adsorbing 0.01 M and 0.05 M solutions of TTF on the magnetic properties of HG. The value of the magnetization drastically decreases on adsorption of TTF and TCNE, although the basic trend in the temperature-variation of magnetization remains the same. Thus, the graphene sample continues to show room-temperature hysteresis. On increasing the concentration of TTF or TCNE, the magnetization value decreases progressively. Interestingly, TTF has a greater effect than TCNE, even though the magnitude of adsorption of TCNE on HG is greater. Magnetic hysteresis of HG persists even after adsorption of TTF and TCNE. The value of $M_S$ at 300 K decreases on adsorption of TTF and TCNE, the decrease being larger in the case of former. The Curie-Weiss temperature, $\theta_p$, also decreases markedly on adsorption of these molecules. Thus, $\theta_p$ of HG becomes -485 K and -83 K after adsorbing 0.05 M TCNE and TTF respectively (compared to -3340 K for pure HG). Clearly, charge-transfer interaction between graphene and TTF (TCNE)[12] is responsible for affecting the magnetic properties. In the literature, there is some evidence to show that adsorption of $H_2O$[22] and interaction of acids[23] reduce the magnetization of nanographite. Potassium clusters also reduce the magnetization of nanographite.[24] In these cases, the reduction in magnetization has been interpreted as due to the interaction with lonepair orbitals as well as charge-transfer with graphene sheets.

In conclusion, graphene samples with the number of layers in the 2-7 range, prepared by different methods, show prominent ferromagnetic features along with the antiferromagnetic characteristics. The value of magnetization as well as the other magnetic properties vary from sample to sample. Considering that ferromagnetism is primarily the result of edge effects, it is not surprising that HG shows the best FM features. Besides the smallest number of layers, this sample has the smallest flakes and therefore more edges. Interestingly, the intensity ratio of the D- band with respect to the G-band in the Raman spectrum is higher in HG than in EG. It is noteworthy that preliminary studies show that the EG-H sample, probably containing mainly single layer graphene, shows much larger magnetization than HG.[21] At present, we are investigating EG-H samples as well as other samples containing primarily single-layer graphene.

**Acknowledgement**: The authors thank Prof. S. V. Bhat, Indian Institute of Science, for EPR measurements.

**Figure captions:**

Figure 1: AFM images and the associated height profiles and Raman spectra of (a) HG and (b) EG samples.

Figure 2: Temperature variation of magnetization of EG, DG and HG at 500 Oe showing the ZFC and FC data. The insets show the magnetization data at 1 T.

Figure 3: Magnetic hysteresis in EG, DG and HG at 300 K. Inset shows magnetic hysteresis in DG at 5 K.

Figure 4: (a) Comparison of the magnetic properties of EG, DG and HG. (b) Variation of the number of layers and sample area.

Figure 5: Temperature-variation of the magnetization of HG samples (500 Oe) after adsorption of 0.01 M and 0.05 M TTF solutions. The magnetization data given in the figure are corrected for the weight of adsorbed TTF. Magnetic hysteresis data at 300 K are shown as insets. In the case of 0.05 M TTF-HG, magnetization data are shown at 1 T as an inset. Magnetization data of HG with adsorbed TCNE are similar to those with TTF, except that the decrease in magnetization relative to pure HG is smaller.

Table 1 Magnetic properties of graphene samples

| Sample | M (emu/g) (FC) | | $|\theta_p|$ | Ms (T=5 K) | $M_R$ (T=300 K) |
| --- | --- | --- | --- | --- | --- |
| | T=5 K (H=1T) | T=300 K (H=0.3 T) | | | |
| HG | 0.527 | 0.393 | 3335 | 0.69 | 0.053 |
| DG | 0.281 | 0.127 | 686 | 0.45 | 0.0162 |
| EG | 0.258 | 0.007 | 230 | 0.062 | 0.0032 |

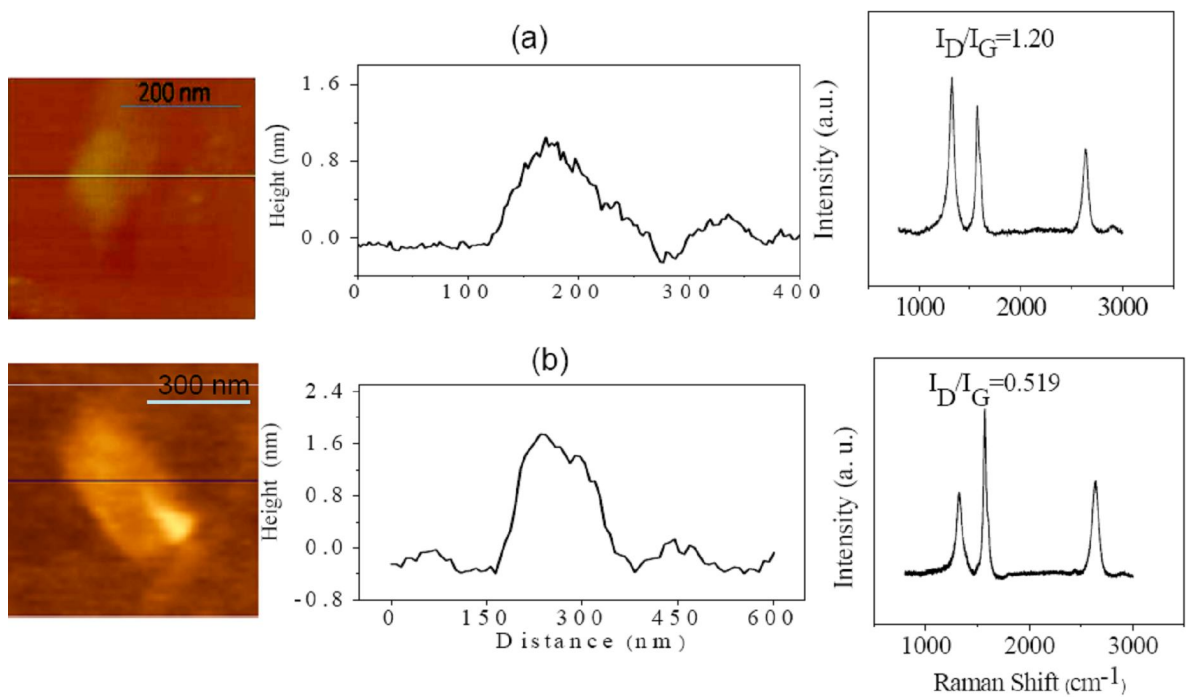

Figure 1

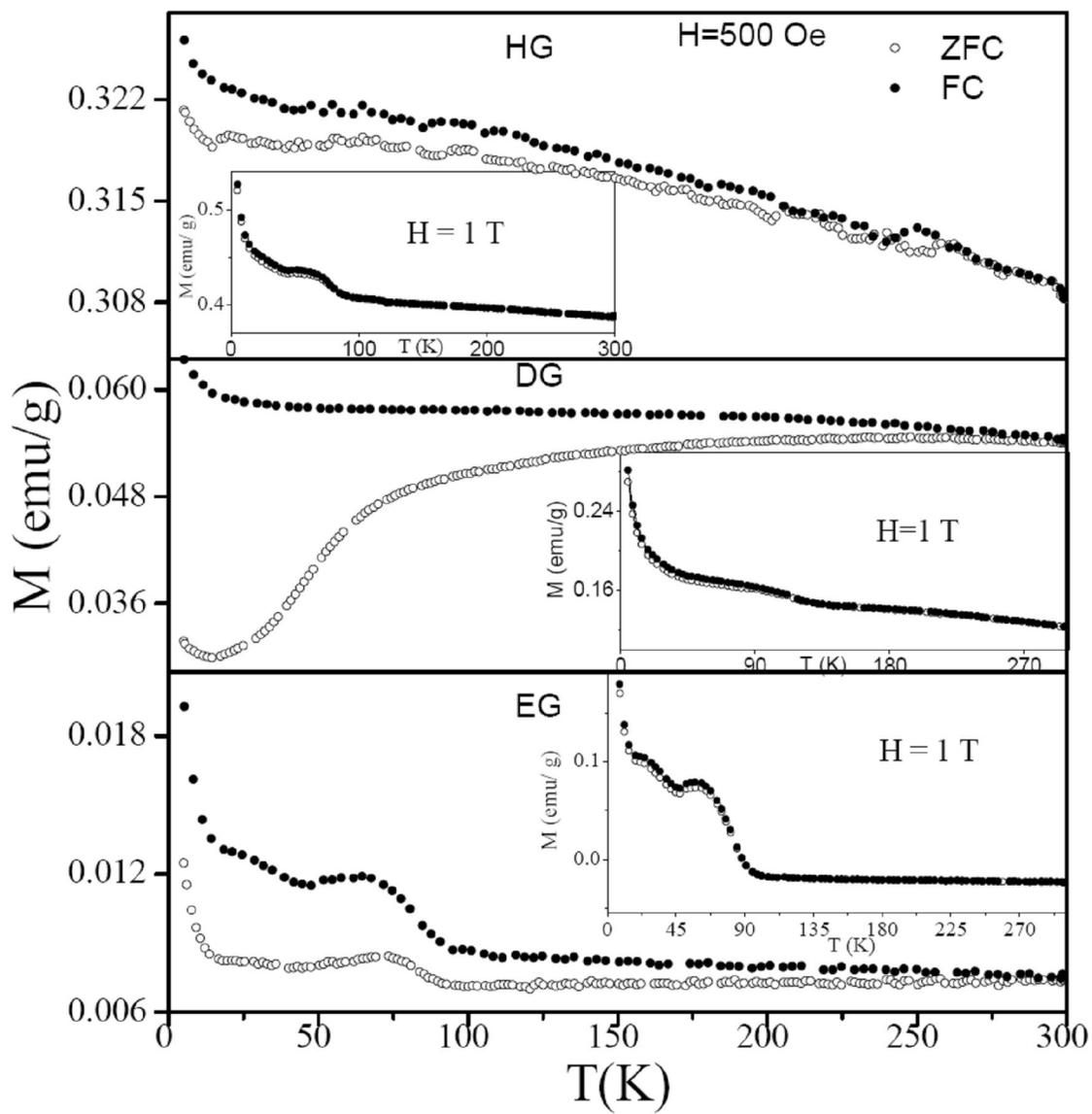

Figure 2

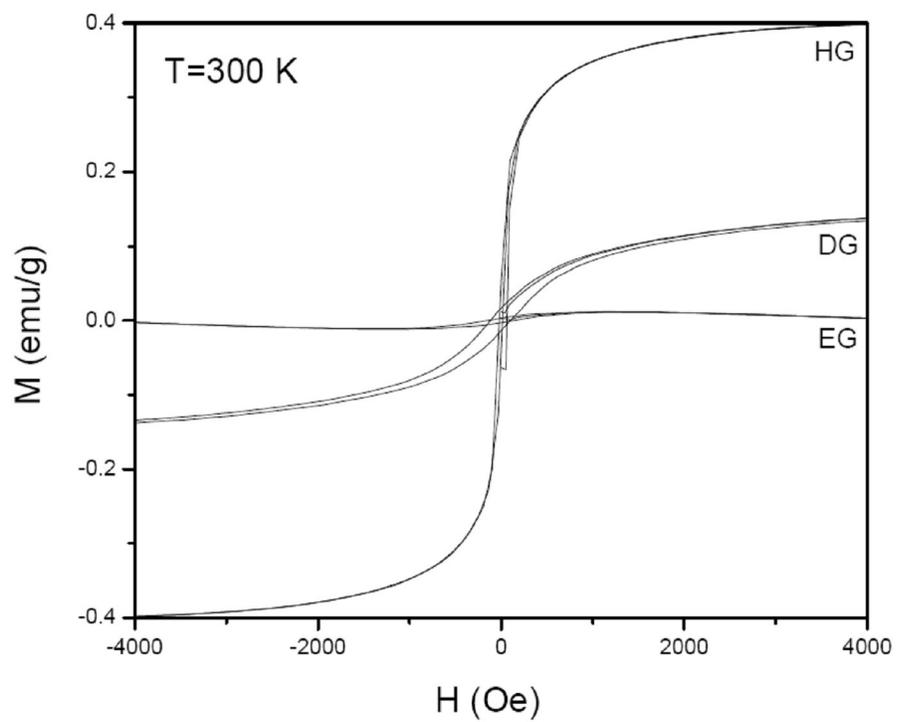

Figure 3

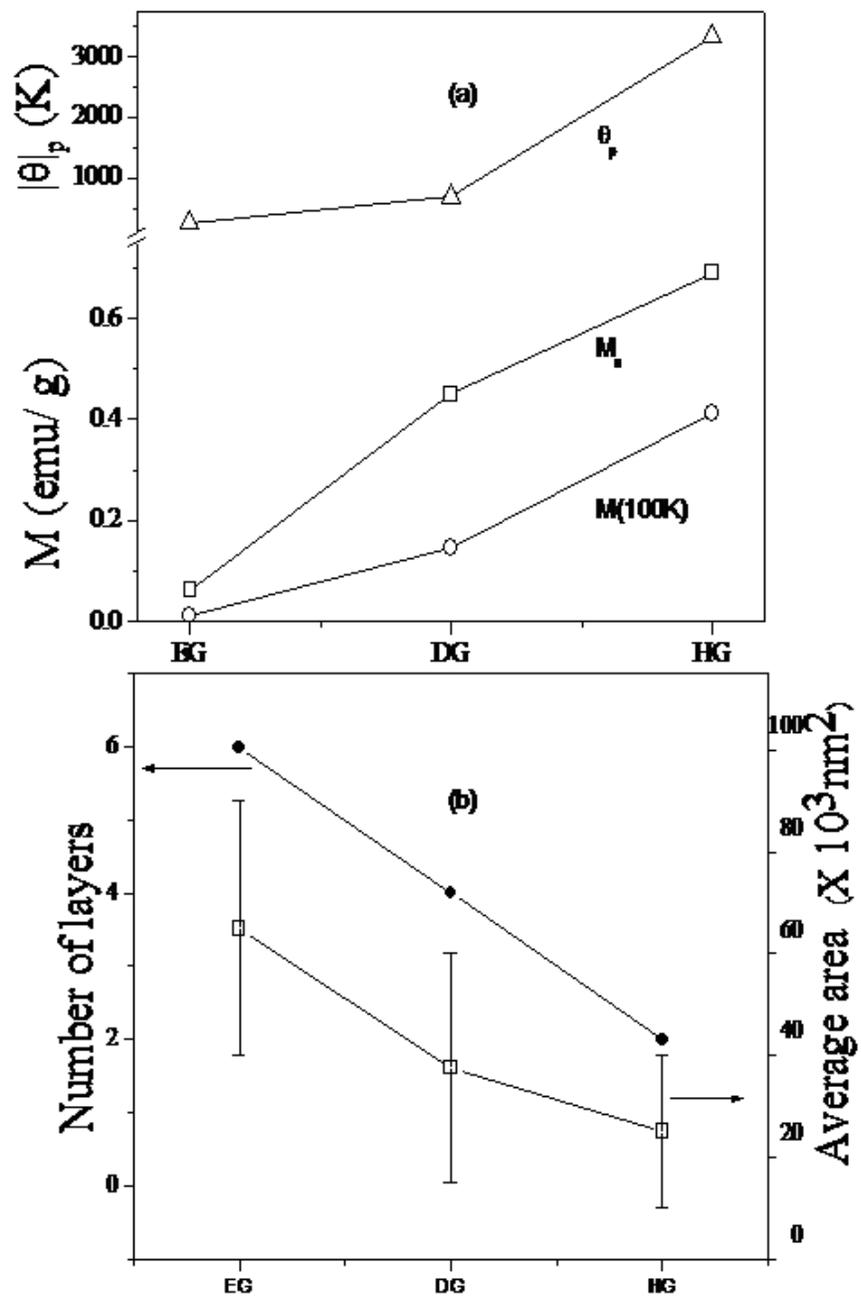

Figure 4

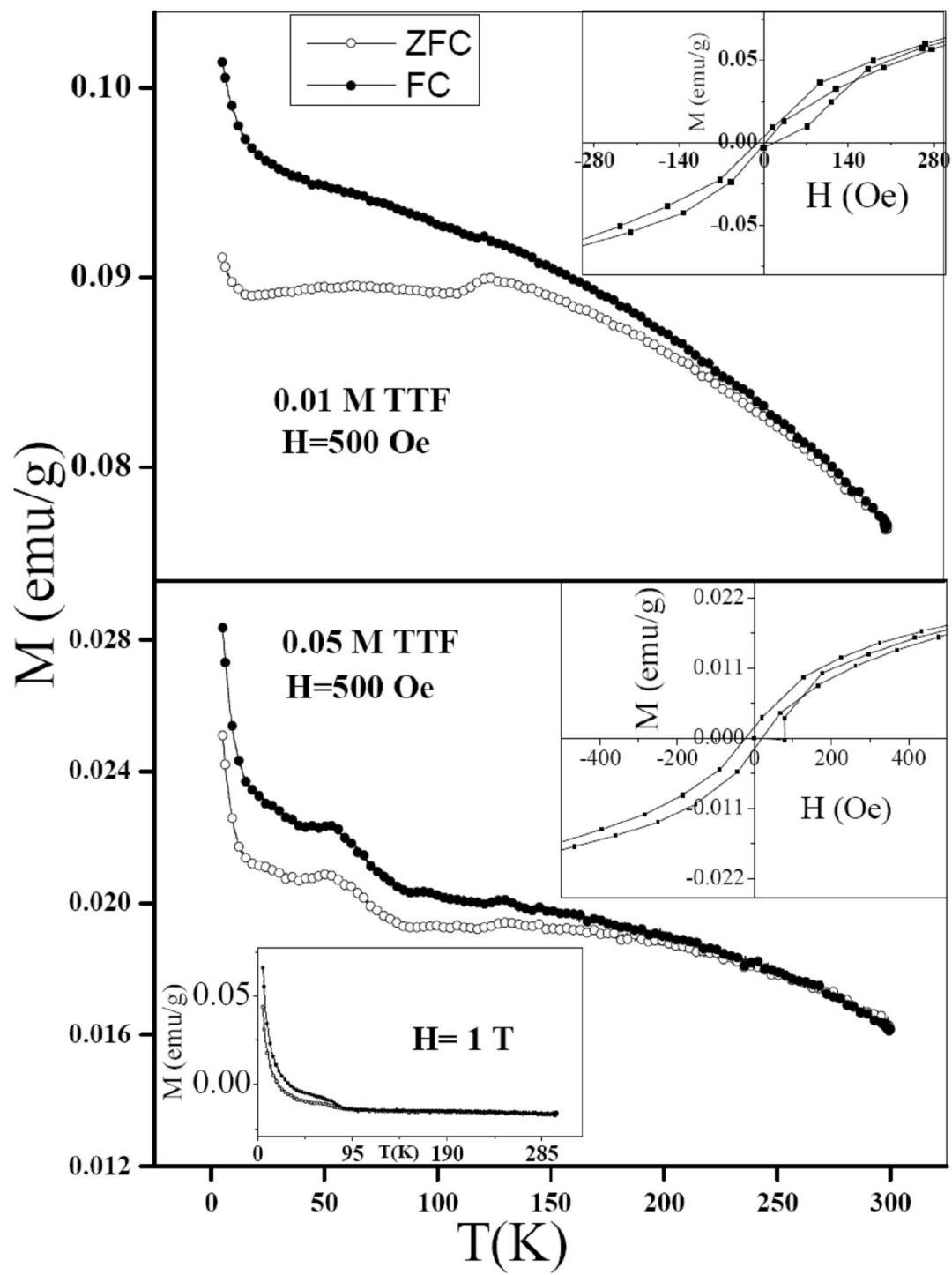

Figure 5